# A Silicon Photonic Transmitter for High Energy Physics and Cryogenic Detectors

A. Quinn[1], S. Moazeni[2]

1. Fermi National Accelerator Laboratory, Kirk Rd & Pine St, 60510, Batavia, USA

2. University of Washington, 1410 NE Campus Parkway, 98195, Seattle, USA

*aquinn@fnal.gov*

***Abstract*—We demonstrate a silicon photonic transmitter designed for reading out physics detectors in extreme environments. The transmitter operates at room and cryogenic temperature and consists of a micro-ring modulator and a co-designed CMOS serializer and driver.**

## I. Introduction

Large pixel detector systems in high energy physics (HEP) can generate terabytes of data per second (TB/s), which must be read out and transferred to a counting room for analysis and data processing [1]. The readout challenge is complicated by the fact that detectors are often located in extreme environments like cryogenic, radiation-hard, or electromagnetic (EM) sensitive environments, while processing can take place meters to miles away.

Optical links enabled by integrated silicon photonic circuits are a crucial technology for meeting future detector requirements due to their extremely high bandwidth readout with low power, low heat load for cryogenic detectors, and resistance to noise and crosstalk [2].

Silicon photonic modulators are extremely widespread in commercial data centers, but their adoption in scientific detectors requires carefully integrating the silicon photonics with custom front end electronics and validating the operation of the entire electronic-photonic system in extreme environments.

We implement a prototype of a silicon photonic transmitter combining a cryogenic detector readout integrated circuit (ROIC) with a photonic integrated circuit (PIC). Section II describes the components of this transmitter, and Section III describes our approach to simple, low-cost packaging of the link for testing at room temperature and at 100 Kelvin. By leveraging IP from the high energy physics community and testing our link at cryogenic temperature, this work is a significant step toward the implementation of silicon photonic links in a practical detector.

## II. Transmitter Design

The transmitter consists of two integrated circuits: the University of Washington Photonic Integrated Circuit V2 (UW-PIC2) and the Skipper-CCD Parallel Readout Circuit V3 (SPROCKET3).

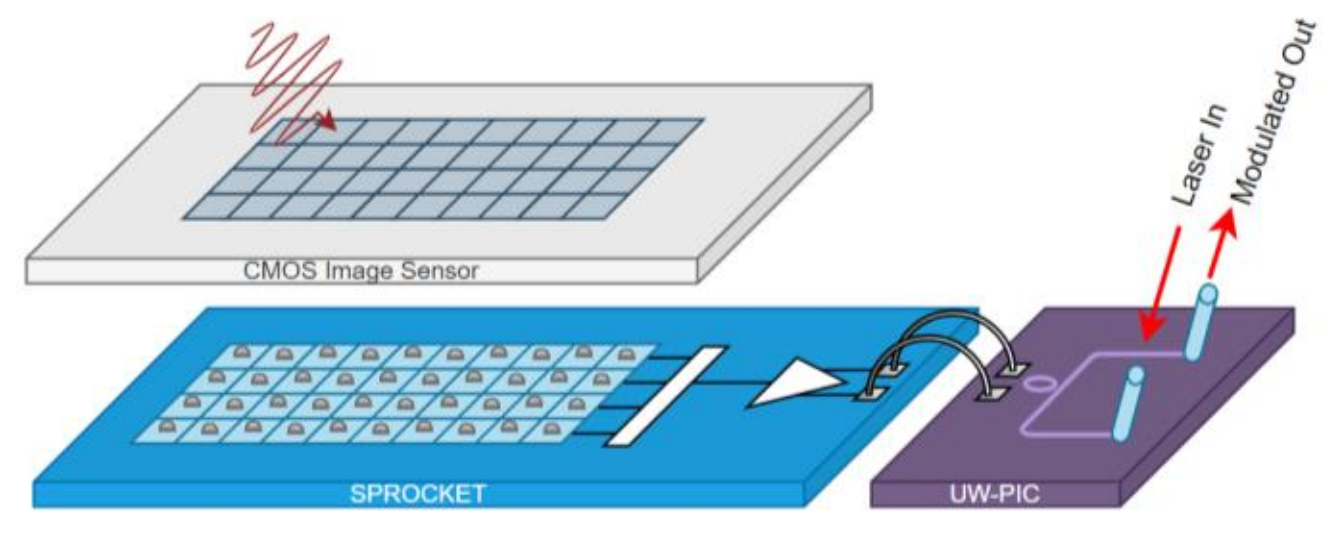

Fig. 1. Conceptual diagram of a photonic transmitter based on SPROCKET3 and UW-PIC2, combined with a CMOS image sensor.

UW-PIC2 is a photonic integrated circuit with sixteen micro-ring modulators (MRMs), with a group of four MRMs coupled to each waveguide to enable future wavelength-division multiplexing. Grating couplers are used to couple light into the chip, and two loopback paths are included for characterizing insertion loss. Each MRM has a dedicated resistive heater for thermal tuning.

SPROCKET3 is a pixelated integrated circuit designed for high-bandwidth hybrid readout of Skipper-in-CMOS sensors. The chip is designed to be vertically integrated with a Skipper-in-CMOS monolithic active pixel sensor (MAPS), and is intended to operate at 100 degrees Kelvin to achieve ideal noise performance from the sensor. Each SPROCKET3 pixel contains a dedicated preamplifier, analog pile-up circuit, and ADC based on a novel compact serial SAR architecture. The front end architecture remains substantially identical to SPROCKET2 [3], but SPROCKET3 scales up the design from a single test pixel to an array of 20,000 pixels, along with array-level bias, control, and readout circuitry.

To serialize data from the pixel array for off-chip transmission, SPROCKET3 implements a serializer derived from the low-power Gigabit Transceiver (lpGBT) developed at CERN. The lpGBT is radiation-tolerant 10.24 Gb/s transmitter developed in 65nm CMOS for the Phase 2 upgrade of the Large Hadron Collider [4]. In SPROCKET3, an additional custom serialization stage is added to increase the maximum data rate to 20.48 Gb/s. The lpGBT's supply voltage of 1.2V is too low to achieve satisfactory modulation of the micro-ring modulators, so we implement a custom AC-coupled differential driver with the DC bias point chosen to result in a swing of 0 to 2.4V at the MRM input.

SPROCKET3 and UW-PIC2 are placed adjacent to one another on a single PCB which provides power and slow control. A single micro-ring modulator on UW-PIC2 is selected to be wirebonded directly to the driver on SPROCKET3. UW-PIC2 is interfaced to an optical fiber by aligning an Oz Optics v-groove assembly with the grating couplers on the surface of the chip.

## III. Packaging and Test

In order to achieve repeatable measurement results, the v-groove assembly must be positioned with sub-micrometer accuracy relative to the surface of the PIC. For this purpose, we used a custom-built 5-axis positioner combining an Alessi REL-6100 probe station, a ThorLabs PY005 Pitch-Yaw stage, and a machined metal arm to hold the v-groove assembly. Initial coarse alignment was carried out visually using the microscope built into the Alessi probe station, while fine alignment was carried out by gradient descent using optical return loss. After achieving satisfactory alignment, we secured the v-groove assembly to the surface of the PIC with NOA 86TLH UV-curable epoxy. A 3D-printed structure was added to provide additional mechanical support to the fibers.

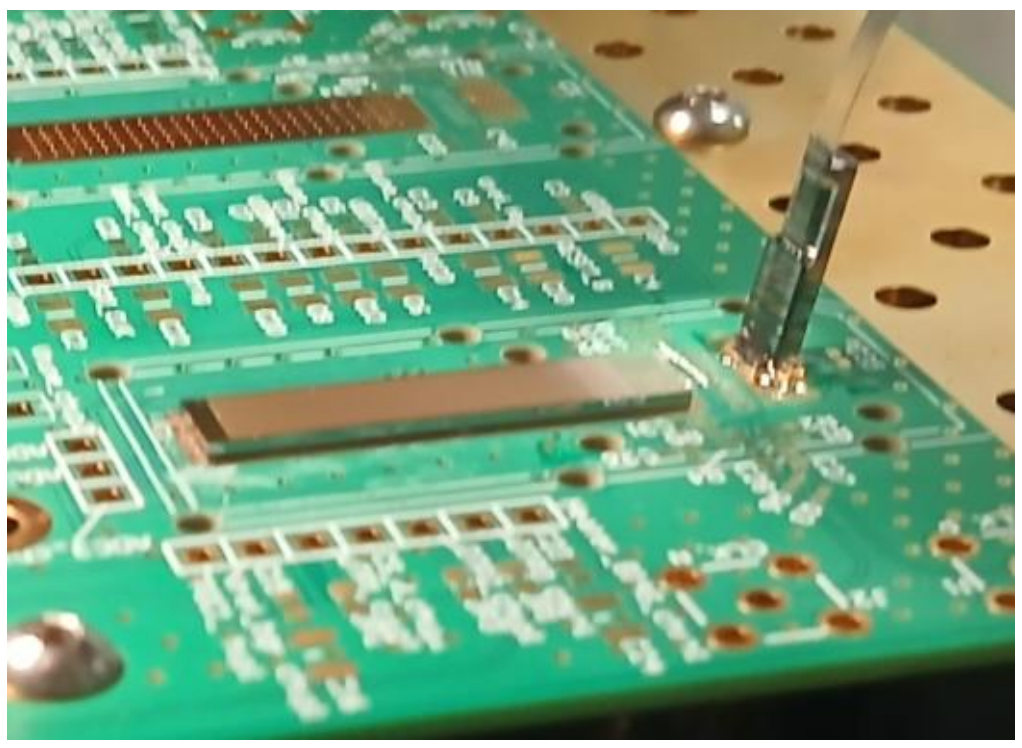

Fig. 2. SPROCKET3 and UW-PIC2 mounted on a PCB, with v-groove assembly (top right), before adding 3D-printed supports.

To test the link, we injected laser light in the 1300 to 1310 nm range, tuning the polarization of the light mechanically to achieve maximum return power. For a receiver, we used a ThorLabs PM100USB optical power meter for low-speed measurements and a Keysight N1092A optical sampling scope for high-speed measurements. Power, bias voltages, and digital slow control were provided by the Spacely-Caribou open source test hardware platform [5].

We tested the device both at room temperature and in a closed-cycle cryostat achieving a base temperature of 66 degrees Kelvin. Initial loopback tests have been performed to verify that both integrated circuits are functional at 66K. Thermal tuning of the link was successful at both temperatures as shown in Fig. 3, with a tuning slope of 393 pm/V at room temperature and 288 pm/V at 66 Kelvin. More extensive characterization of the link performance including its modulation amplitude and bandwidth is underway with results to be presented at the conference.

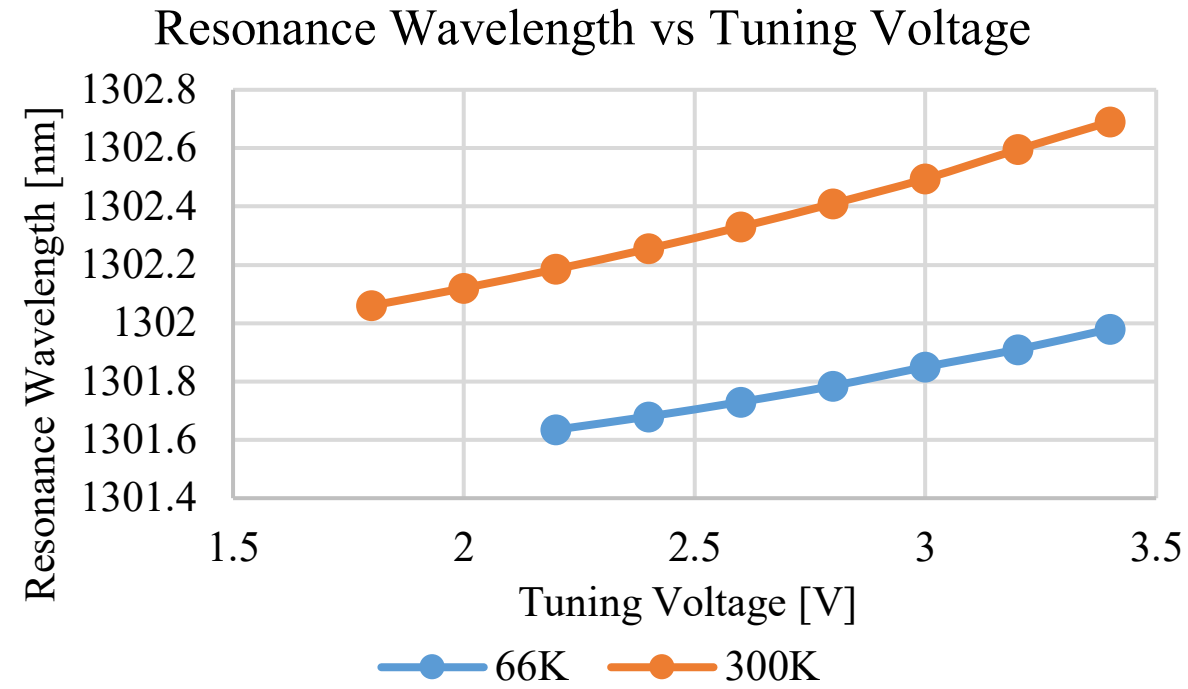

Fig. 3. Measurement results of the wavelength of the selected micro-ring modulator versus the voltage applied to the resistive heater.

## IV. Conclusions

We present the design of a prototype silicon photonic transmitter optimized for use cases in high energy physics, together with an approach to packaging and preliminary test results at room and cryogenic temperature. Future work includes full characterization of the link performance at cryogenic temperature, as well as under irradiation.

## Acknowledgments

Thanks to the EMIT Lab at University of Washington for providing UW-PIC2 and support with packaging. Thanks to Sergey Los at Fermilab for designing the 3D-printed supports and providing many useful suggestions.